\documentclass[preprint,prd,aps,showpacs,showkeys,nofootinbib]{revtex4}
\usepackage{graphicx}
\usepackage{dcolumn}
\usepackage{bm}
\usepackage{color}
\usepackage[dvipsnames]{xcolor}
\usepackage{amssymb}
\usepackage{epstopdf}

\definecolor{black-blue}{RGB}{77,116,175}
\definecolor{black-yellow}{RGB}{231,162,33}
\definecolor{black-green}{RGB}{144,180,58}
\definecolor{black-red}{RGB}{246,95,50}

\begin{document}

\title{The top quark rare decays with flavor violation}
\author{Ming-Yue Liu$^{1,2,3}$, Shu-Min Zhao$^{1,2,3}$\footnote{zhaosm@hbu.edu.cn}, Song Gao$^{1,2,3}$, Xing-Yu Han$^{1,2,3}$, Tai-Fu Feng$^{1,2,3,4}$.}

\affiliation{$^1$ Department of Physics, Hebei University, Baoding 071002, China}
\affiliation{$^2$ Hebei Key Laboratory of High-precision Computation and Application of Quantum Field Theory, Baoding, 071002, China}
\affiliation{$^3$ Hebei Research Center of the Basic Discipline for Computational Physics, Baoding, 071002, China}
\affiliation{$^4$ Department of Physics, Chongqing University, Chongqing 401331, China}
\date{\today}

\begin{abstract}
In the present study, we  investigate the decays of the top quark $t\rightarrow c\gamma$, $t\rightarrow cg$, $t\rightarrow cZ$ and $t\rightarrow ch$. They are extremely rare processes in the Standard Model (SM). As the $U(1)$ extension of the minimal supersymmetric standard model (MSSM), the $U(1)_X$SSM features new superfields such as the right-handed neutrinos and three Higgs singlets. We analyze the effects of different sensitive parameters on the results and make reasonable theorecial predictions, which provides a useful reference for future experimental development. Considering the constraint of the updated experimental data, the numerical results show that the branching ratios of all four processes $t\rightarrow c\gamma,~cg,~cZ,~ch$ can reach the same order of magnitude as their experimental upper limits. Among them, $\tan\beta$ has the most obvious effect on each process and is the main parameter. $g_X$, $g_{YX}$, $\mu$, $M_2$, $\lambda_H$, $M_{U23}^2$ and $M_{Q23}^2$ are important parameters for the processes, and have effects on the numerical results.
\end{abstract}


\maketitle

\section{Introduction}
The top quark was discovered in 1995 by a team of D0 and CDF experiments located at the Fermi National Accelerator Laboratory (Fermilab) in the United States \cite{j1,j2}. This discovery is important for the validation of the SM and the study of particle physics. The study of the nature and behavior of the top quark helps us to understand physical processes such as the origin of the mass of elementary particles, weak and strong interactions. The existence and properties of the top quark have been verified several times in experiments, including the D0 and CDF experiments at Fermilab and the ATLAS \cite{j3} and CMS \cite{j4} experiments at the Large Hadron Collider (LHC) in Geneva, Switzerland. These experiments investigate aspects of the nature, decay modes and interactions of the top quark through high-energy collision and particle detection techniques. The next generation of LHC will produce top quarks in large quantities. At the upgraded Fermilab, an integrated luminosity of $10 fb^{-1}$ will produce about $8\times{10^4}$ top quarks, while at the same luminosity the LHC will produce about 100 times as many \cite{1,2,3}.

Top quark decays with flavor violation refer to the decay processes of the top quark that violate the conservation of flavor, specifically the violation of lepton or quark flavor. While the SM predicts that the top quark predominantly decays into a $W$ boson and a bottom quark, extensions beyond the SM allow for additional decay modes that involve different quarks or leptons \cite{j41,j42}. It is worth noting that specific details about the nature and extent of flavor violation during top quark decays require more in-depth analysis. The study of heavy particle decays via flavor-changing neutral-currents (FCNC) has played an important role in testing the SM and exploring new physics beyond the SM \cite{j5,j6,j61}. In the SM branching ratios of the FCNC of the top quarks $t\rightarrow c\gamma$, $t\rightarrow cg$, $t\rightarrow cZ$, $t\rightarrow ch$ are highly suppressed, and beyond the detection capabilities of the LHC in the near future \cite{j7,j8,j9}. However, exotic mechanisms from new physics can greatly increase these branching ratios of \cite{4}, which can be detected in the future.
The SM predictions \cite{22} and the latest upper bounds on the branching ratios of $t\rightarrow c\gamma$, $t\rightarrow cg$, $t\rightarrow cZ$ and $t\rightarrow ch$ at the 95\% confidence level (C.L.) \cite{5} are given in Table \ref {1}. We can find that the current experimental bounds are much higher than the predictions of SM.

As the heaviest elementary particle in the SM with mass on the electroweak scale, the top quark is likely to be more sensitive to new physics. Kinematically, it can reach many FCNC decay modes such as $t\rightarrow c\gamma$, $t\rightarrow cg$, $t\rightarrow cZ$ and $t\rightarrow ch$, where $h$ is the lightest CP-even Higgs boson. In the SM, these FCNC decay modes are highly suppressed by the GIM mechanism, with branching ratios typically of the order of $10^{-15}-10^{-12}$ \cite{1,2,3,6,7,8,9,10,11,12}, a relatively small order of magnitude. On the other hand, the observation of any such FCNC top-quark decay would be strong evidence of new physics. Therefore, detecting those top-quark rare decays at the LHC provides a good window to search for new physics beyond the SM. We learn some theoretical predictions for the branching ratios of top quark rare decays in new physics extensions, such as in supersymmetric  (SUSY) models with R-parity conservation.
These branching ratios can reach $Br(t\rightarrow c\gamma)\sim 10^{-6}$, $Br(t\rightarrow cg)\sim 10^{-5}$,
$Br(t\rightarrow cZ)\sim 10^{-6}$ \cite{13,14}. The branching ratios from SUSY without R-parity conservation can reach
$Br(t\rightarrow c\gamma)\sim 10^{-6}$, $Br(t\rightarrow cg)\sim 10^{-4}$, $Br(t\rightarrow cZ)\sim 10^{-7}$,
$Br(t\rightarrow ch)\sim 10^{-4}$ \cite{15,16}. The branching ratios in the two Higgs doublet models can be realized up to
$Br(t\rightarrow c\gamma)\sim 10^{-6}$, $Br(t\rightarrow cg)\sim 10^{-4}$,
$Br(t\rightarrow cZ)\sim 10^{-7}$, $Br(t\rightarrow ch)\sim 10^{-3}$ \cite{17,18,19,20}.  In the extension of the minimal supersymmetric standard model with additional local $U(1)_{B-L}$ gauge symmetry (B-LSSM) the branching ratios reach $Br(t\rightarrow c\gamma)\sim 5\times10^{-7}$, $Br(t\rightarrow cg)\sim 2\times10^{-6}$, $Br(t\rightarrow cZ)\sim 4\times10^{-7}$, $Br(t\rightarrow ch)\sim 3\times10^{-9}$ \cite{21}.

\begin{table*}
\caption{ The SM predictions and experimental bounds on the decays $t\rightarrow cV, ch$}
\begin{tabular*}{\textwidth}{@{\extracolsep{\fill}}|l|l|l|l|l|@{}}
\hline
Decay&$Br(t\rightarrow c\gamma)$&$Br(t\rightarrow cg)$&$Br(t\rightarrow cZ)$&$Br(t\rightarrow ch)$\\
\hline
SM&$4.6\times10^{-14}$~~~~~~~&$4.6\times10^{-12}$~~~~~~~&~$1\times10^{-14}$~~~~~~~&$~3\times10^{-15}$~~~~~~~\\
\hline
Upper Limit(95\%C.L.)&$1.8\times10^{-4}$&$2\times10^{-4}$&~$5\times10^{-4}$~&~$1.1\times10^{-3}~~~~~~$\\
\hline
\end{tabular*}
\label{1}
\end{table*}

In this work, we explore top quark decays with flavor violation under the $U(1)_X$SSM. The $U(1)_X$SSM is an extension of the MSSM that incorporates an extra $U(1)_X$ gauge symmetry. Its local gauge group is $SU(3)_C\times SU(2)_L \times U(1)_Y\times U(1)_X$ \cite{27,28,29}. Compared to the MSSM, in the $U(1)_X$SSM
we add three new Higgs singlets  $\hat{\eta},~\hat{\bar{\eta}},~\hat{S}$ and three-generation right-handed neutrinos $\hat{\nu}_i$. The right-handed neutrinos have the function of both generating tiny mass to light neutrinos via a see-saw mechanism and providing a new dark matter candidate light sneutrino. The presence of right-handed neutrinos, sneutrinos and additional Higgs singlets alleviates the so-called small hierarchy problem arising in the MSSM.  MSSM exists $\mu$ problem, while in the $U(1)_X$SSM \cite{30} this problem can be alleviated by the $S$ field after vacuum spontaneous breaking.

The outline of this paper is as follows. In Sec.II, we briefly introduce the essential content of the $U (1)_X$SSM, including its superpotential, the general soft breaking terms, the rotations and interactions of the eigenstates "EWSB". In Sec.III, we provide analytical expressions for the branching ratios of the $t\rightarrow cV, ch$ ($V=\gamma,~Z,~g$) decays in the $U(1)_X$SSM. In Sec.IV, we give the corresponding parameters and numerical analysis. In Sec.V, we present a summary of this article.

\section{The $U(1)_X$SSM}

In this section, we will provide some overview of $U(1)_X$SSM.  $U(1)_X$SSM includes a local gauge group $SU(3)_C\times SU(2)_L \times U(1)_Y$ with the same gauge group as the SM and the MSSM. It is a $U(1)_X$ extension of the MSSM, and has the local gauge group $SU(3)_C\times SU(2)_L \times U(1)_Y\times U(1)_X$ \cite{31,32,33}. In addition to the MSSM, the field spectrum in $U(1)_X$SSM contains new superfields, which are the right-handed neutrinos $\hat{\nu}_i$ and the three Higgs singlets $\hat{\eta},~\hat{\bar{\eta}},~\hat{S}$. Through the seesaw mechanism, the lighter neutrinos gain tiny masses at the tree level. The formation of the $5\times5$ mass-squared matrix is due to the mixing of the neutral CP-even parts of $H_u$, $H_d$, $\eta$, $\bar{\eta}$, and $S$. To obtain the 125.25 GeV Higgs particle mass \cite{34,35}, loop corrections should be considered. These sneutrinos are decomposed into CP-even sneutrinos and CP-odd sneutrinos, and their mass-squared matrices are both expanded to $6\times6$.

The superpotential in $U(1)_X$SSM is denoted by:
\begin{eqnarray}
&&W=l_W\hat{S}+\mu\hat{H}_u\hat{H}_d+M_S\hat{S}\hat{S}-Y_d\hat{d}\hat{q}\hat{H}_d-Y_e\hat{e}\hat{l}\hat{H}_d+\lambda_H\hat{S}\hat{H}_u\hat{H}_d
\nonumber\\&&\hspace{0.6cm}+M_{\hat{\eta}} \hat{\eta}\hat{\bar{\eta}}+\lambda_C\hat{S}\hat{\eta}\hat{\bar{\eta}}+\frac{\kappa}{3}\hat{S}\hat{S}\hat{S}
+Y_u\hat{u}\hat{q}\hat{H}_u+Y_X\hat{\nu}\hat{\bar{\eta}}\hat{\nu}
+Y_\nu\hat{\nu}\hat{l}\hat{H}_u.\label{G1}
\end{eqnarray}

In Eq.(\ref{G1}) the vacuum expectation value of $\hat{\bar{\eta}}$
produces the Majorana mass of the right-handed neutrino
through $Y_X\hat{\nu}\hat{\bar{\eta}}\hat{\nu}$. While the right-handed neutrino mixes with the left-handed neutrino through $Y_\nu\hat{\nu}\hat{l}\hat{H}_u$.

The vacuum expectation values(VEVs) of the Higgs superfields $H_u$, $H_d$, $\eta$, $\bar{\eta}$ and $S$ are
denoted by $v_u$, $v_d$, $v_\eta$, $v_{\bar{\eta}}$ and $v_S$ respectively. Two angles are defined as $\tan\beta=v_u/v_d$
and $\tan\beta_\eta=v_{\bar{\eta}}/v_{\eta}$. The explicit forms of the two Higgs doublets and three Higgs singlets are written as:
\begin{eqnarray}
&&\hspace{0cm}\eta={1\over\sqrt{2}}\Big(v_{\eta}+\phi_{\eta}^0+iP_{\eta}^0\Big),~~~
\bar{\eta}={1\over\sqrt{2}}\Big(v_{\bar{\eta}}+\phi_{\bar{\eta}}^0+iP_{\bar{\eta}}^0\Big),~~
S={1\over\sqrt{2}}\Big(v_{S}+\phi_{S}^0+iP_{S}^0\Big),
\nonumber\\&&H_{u}=\left(\begin{array}{c}H_{u}^+\\{1\over\sqrt{2}}\Big(v_{u}+H_{u}^0+iP_{u}^0\Big)\end{array}\right),
~~~~~~
H_{d}=\left(\begin{array}{c}{1\over\sqrt{2}}\Big(v_{d}+H_{d}^0+iP_{d}^0\Big)\\H_{d}^-\end{array}\right).
\end{eqnarray}

The soft SUSY breaking terms of $U(1)_X$SSM are shown as:
\begin{eqnarray}
&&\mathcal{L}_{soft}=\mathcal{L}_{soft}^{MSSM}-B_SS^2-L_SS-\frac{T_\kappa}{3}S^3-T_{\lambda_C}S\eta\bar{\eta}
+\epsilon_{ij}T_{\lambda_H}SH_d^iH_u^j\nonumber\\&&\hspace{1cm}
-T_X^{IJ}\bar{\eta}\tilde{\nu}_R^{*I}\tilde{\nu}_R^{*J}
+\epsilon_{ij}T^{IJ}_{\nu}H_u^i\tilde{\nu}_R^{I*}\tilde{l}_j^J
-m_{\eta}^2|\eta|^2-m_{\bar{\eta}}^2|\bar{\eta}|^2-m_S^2S^2\nonumber\\&&\hspace{1cm}
-(m_{\tilde{\nu}_R}^2)^{IJ}\tilde{\nu}_R^{I*}\tilde{\nu}_R^{J}
-\frac{1}{2}\Big(M_S\lambda^2_{\tilde{X}}+2M_{BB^\prime}\lambda_{\tilde{B}}\lambda_{\tilde{X}}\Big)+h.c~.
\end{eqnarray}
$\mathcal{L}_{soft}^{MSSM}$ are soft breaking terms of MSSM. The particle content and charge assignments for $U(1)_X$SSM are shown in the Table \ref {JJ1}.
In our previous work, we have shown that the $U(1)_X$SSM is anomaly free \cite{32}. In the $U(1)_X$SSM, $U(1)_Y$ and $U(1)_X$ are two Abelian groups. We denote the $U(1)_Y$ charge by $Y^Y$. The $U(1)_X$ charge by $Y^X$, and the presence of these two Abelian groups gives rise to a new effect that is not found in the MSSM or other SUSY models with only one Abelian gauge group: the gauge kinetic mixing.

\begin{table}[h]
\caption{ The superfields in $U(1)_X$SSM}
\begin{tabular}{|c|c|c|c|c|c|c|c|c|c|c|c|}
\hline
Superfields & $\hspace{0.1cm}\hat{q}_i\hspace{0.1cm}$ & $\hat{u}^c_i$ & $\hspace{0.2cm}\hat{d}^c_i\hspace{0.2cm}$ & $\hat{l}_i$ & $\hspace{0.2cm}\hat{e}^c_i\hspace{0.2cm}$ & $\hat{\nu}_i$ & $\hspace{0.1cm}\hat{H}_u\hspace{0.1cm}$ & $\hat{H}_d$ & $\hspace{0.2cm}\hat{\eta}\hspace{0.2cm}$ & $\hspace{0.2cm}\hat{\bar{\eta}}\hspace{0.2cm}$ & $\hspace{0.2cm}\hat{S}\hspace{0.2cm}$ \\
\hline
$SU(3)_C$ & 3 & $\bar{3}$ & $\bar{3}$ & 1 & 1 & 1 & 1 & 1 & 1 & 1 & 1  \\
\hline
$SU(2)_L$ & 2 & 1 & 1 & 2 & 1 & 1 & 2 & 2 & 1 & 1 & 1  \\
\hline
$U(1)_Y$ & 1/6 & -2/3 & 1/3 & -1/2 & 1 & 0 & 1/2 & -1/2 & 0 & 0 & 0  \\
\hline
$U(1)_X$ & 0 & -1/2 & 1/2 & 0 & 1/2 & -1/2 & 1/2 & -1/2 & -1 & 1 & 0  \\
\hline
\end{tabular}
\label{JJ1}
\end{table}

This effect can be caused by RGEs. $A_\mu^{'Y}$ and $A_\mu^{'X}$ denote the gauge fields of $U(1)_Y$ and $U(1)_X$ respectively. The form of the covariant derivative of the $U(1)_X$SSM can be written as:
\begin{eqnarray}
&&D_\mu=\partial_\mu-i\left(\begin{array}{cc}Y^Y,&Y^X\end{array}\right)
\left(\begin{array}{cc}g_{Y},&g{'}_{{YX}}\\g{'}_{{XY}},&g{'}_{{X}}\end{array}\right)
\left(\begin{array}{c}A_{\mu}^{\prime Y} \\ A_{\mu}^{\prime X}\end{array}\right)\;.
\end{eqnarray}

We redefine the following\cite{36,37}:
\begin{eqnarray}
&&\left(\begin{array}{cc}g_{Y},&g{'}_{YX}\\g{'}_{XY},&g{'}_{X}\end{array}\right)
R^T=\left(\begin{array}{cc}g_{1},&g_{YX}\\0,&g_{X}\end{array}\right)~,~~~~
R\left(\begin{array}{c}A_{\mu}^{\prime Y} \\ A_{\mu}^{\prime X}\end{array}\right)
=\left(\begin{array}{c}A_{\mu}^{Y} \\ A_{\mu}^{X}\end{array}\right)\;.
\end{eqnarray}
Finally, the gauge derivative of $U(1)_X$SSM is transformed into:
\begin{eqnarray}
&&D_\mu=\partial_\mu-i\left(\begin{array}{cc}Y^Y,&Y^X\end{array}\right)
\left(\begin{array}{cc}g_{1},&g_{{YX}}\\0,&g_{{X}}\end{array}\right)
\left(\begin{array}{c}A_{\mu}^{Y} \\ A_{\mu}^{X}\end{array}\right)\;.
\end{eqnarray}
The $g_X$ appearing above is the gauge coupling constant for the $U(1)_X$ group. $g_{YX}$ is the mixed gauge coupling constant for the $U(1)_Y$ group and the $U(1)_X$ group.

In the $U(1)_X$SSM, the gauge bosons $A_\mu^{'Y}$ , $A_\mu^{'X}$ and $V_\mu^3$ are mixed together at the tree level.
The mass matrix of gauge bosons can be found in reference \cite{32}.
We use two mixing angles $\theta_{W}$ and $\theta_{W}'$ to obtain the mass eigenvalues of the matrix. $\theta_{W}$ is the Weinberg angle and $\theta_{W}'$ is the new mixing angle. We define $v=\sqrt{v_u^2+v_d^2}$ and $\xi=\sqrt{v_\eta^2+v_{\bar{\eta}}^2}$. The new mixing angle is defined as:
\begin{eqnarray}
\sin^2\theta_{W}'\!=\!\frac{1}{2}\!-\!\frac{[(g_{{YX}}+g_{X})^2-g_{1}^2-g_{2}^2]v^2+
4g_{X}^2\xi^2}{2\sqrt{[(g_{{YX}}+g_{X})^2+g_{1}^2+g_{2}^2]^2v^4+8g_{X}^2[(g_{{YX}}+g_{X})^2-g_{1}^2-g_{2}^2]v^2\xi^2+16g_{X}^4\xi^4}}.
\end{eqnarray}
We derive the eigenvalues of the mass-squared matrix of the
neutral gauge bosons. One is the zero mass corresponding to the photon and the other two values are $Z$ and $Z'$ ,
\begin{eqnarray}
&&m_\gamma^2=0,
\nonumber\\&&m^2_{Z,Z'}=\!\frac{1}{8}\!\Big((g_1^2+g_2^2+(g_{YX}+g_X)^2)v^2+4g_X^2\xi^2\Big)
\nonumber\\&&\mp\sqrt{(g_1^2+g_2^2+(g_{YX}+g_X)^2)^2v^4+8((g_{YX}+g_X)^2-g_1^2-g_2^2)g_X^2v^2\xi^2+16g_X^2\xi^4}.
\end{eqnarray}
The mass matrix for chargino is:
\begin{eqnarray}
m_{\tilde{\chi}^-} = \left(
\begin{array}{cc}
M_2&\frac{1}{\sqrt{2}}g_2v_\mu\\
\frac{1}{\sqrt{2}}g_2v_d&\frac{1}{\sqrt{2}}\lambda_{H} v_S+\mu\end{array}
\right).\label{Y2}
 \end{eqnarray}
This matrix is diagonalized by U and V:\begin{eqnarray}
U^{*} m_{\tilde{\chi}^-} V^{\dagger}= m_{\tilde{\chi}^-}^{diag}.
 \end{eqnarray}
The mass matrix for neutrino is:
\begin{eqnarray}
m_{\nu} = \left(
\begin{array}{cc}
0&\frac{1}{\sqrt{2}}v_uY^T_\nu\\
\frac{1}{\sqrt{2}}v_uY_\nu&\ \sqrt{2}v_{\bar{\eta}}Y_X\end{array}
\right).\label{Y2}
\end{eqnarray}
This matrix is diagonalized by $U^V$:
\begin{eqnarray}
U^{V,*}m_\nu U^{V,\dag} = m_\nu^{dia}.
\end{eqnarray}

In addition, a number of other mass matrices are required in the calculations, all of which can be found in Refs. \cite{32,38}.

\section{ANALYTICAL FORMULA}

\begin{figure}[ht]
\setlength{\unitlength}{5.0mm}
\centering
\includegraphics[width=6.5in]{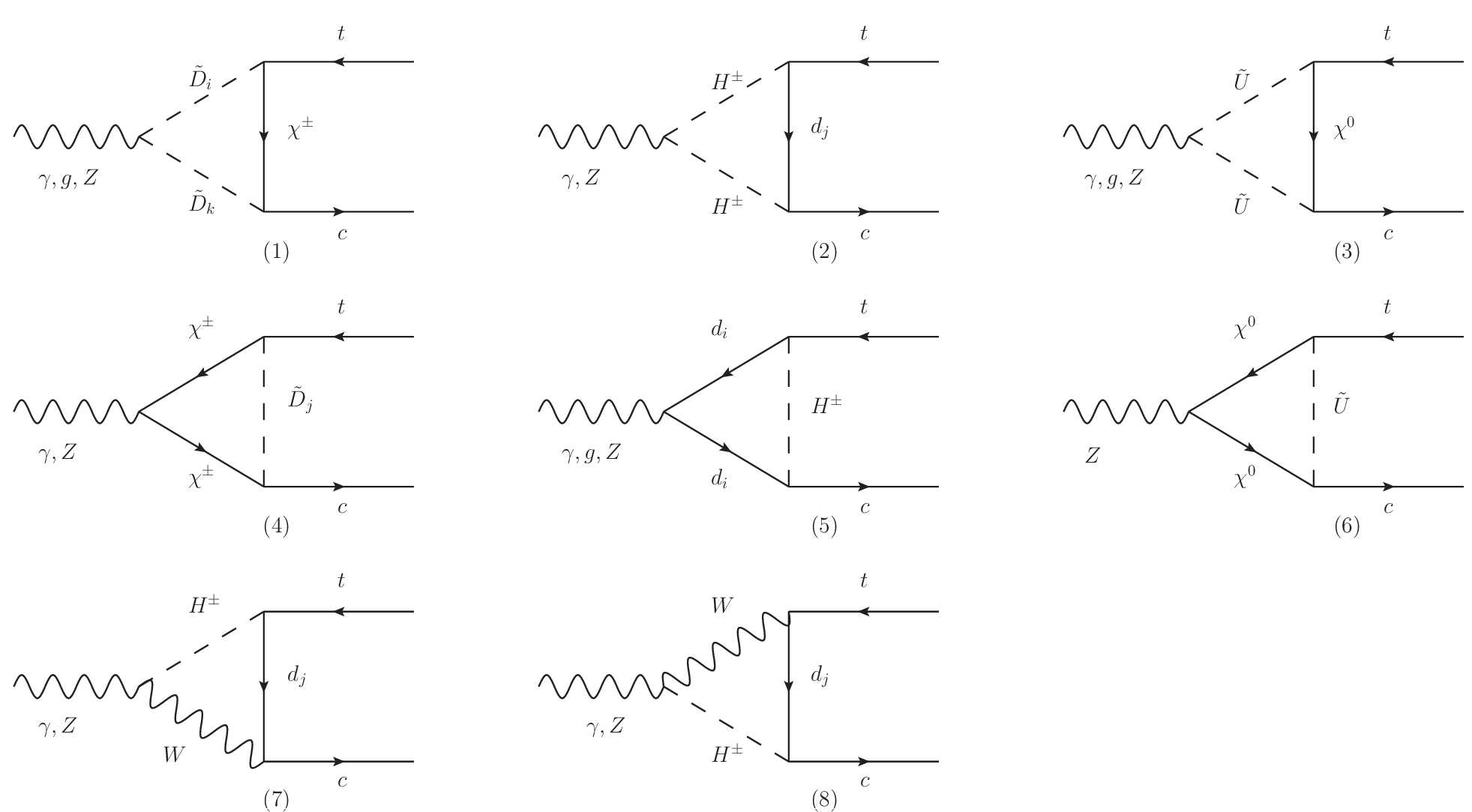}
\caption{Feynman diagrams for the $t\rightarrow c\gamma$, $t\rightarrow cg$, $t\rightarrow cZ$ processes in the $U(1)_X$SSM.}\label{A1}
\end{figure}

\begin{figure}[ht]
\setlength{\unitlength}{5.0mm}
\centering
\includegraphics[width=6.5in]{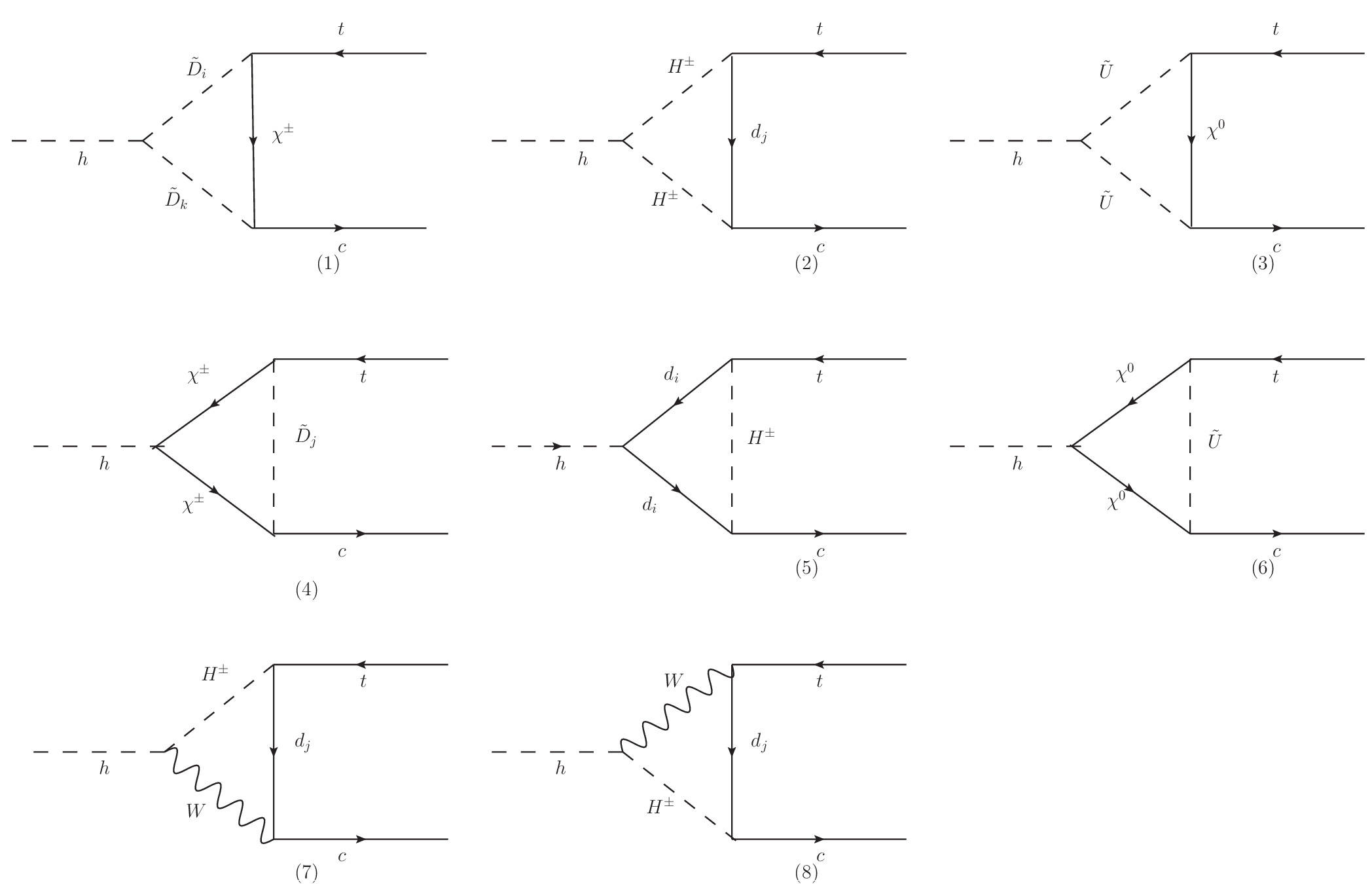}
\caption{Feynman diagrams for the $t\rightarrow ch$ process in the $U(1)_X$SSM.}\label{A2}
\end{figure}

In this section, we focus on the theoretical study of  the top quark processes  $t\rightarrow c\gamma$, $t\rightarrow cg$, $t\rightarrow cZ$ and $t\rightarrow ch$  with flavor violation under the $U(1)_X$SSM. The relevant Feynman diagrams contributing to $t\rightarrow c\gamma$, $t\rightarrow cg$, $t\rightarrow cZ$ and $t\rightarrow ch$ in the $U(1)_X$SSM are presented in Fig.\ref{A1} and Fig.\ref{A2}.

In the $U(1)_X$SSM, the flavor violating amplitude corresponding to the decay process $t\rightarrow cV$ ($V=\gamma, Z$) is written as:
\begin{eqnarray}
&&\mathcal{M}_{t\rightarrow cV}=\varepsilon^\mu\bar{u}_c(p')\Big(A_V\gamma_\mu P_L+iB_V\sigma_{\mu\nu}q^\nu P_L+(L\rightarrow R)\Big)u_t(p).
\end{eqnarray}
In order to better explain how the calculation of the above equation as well as the Feynman diagram in Fig.\ref{A1} is done, we take the calculation of Fig.\ref{A1}(1) as an example. The corresponding amplitude can be written as:
\begin{eqnarray}
&&\mathcal{M}_{t\rightarrow cV}=\sum_{i,j,k,n}\varepsilon^\mu\bar{u}_c(p')\int\frac{d^Dk}{(2\pi)^D}i(A_{\bar{c}\tilde{D}_k\chi^\pm_n }^LP_L+A_{\bar{c}\tilde{D}_k\chi_n^\pm}^RP_R)\frac{i}{p\!\!\!\slash-k\!\!\!\slash-m_{\chi_n^\pm}}\nonumber\\&&
i(A_{\bar{\chi}_n^\pm\tilde{D}_it}^LP_L+A_{\bar{\chi}_n^\pm\tilde{D}_it}^RP_R)\frac{i}{k^2-m_{\tilde{D}_i}^2}iB_{V\tilde{D}_i\tilde{D}_k}(-2k_\mu+q_\mu)
\frac{i}{(k-q)^2-m_{\tilde{D}_k}^2}u_t(p).\label{K1}
\end{eqnarray}

Here, $\varepsilon_\mu$ denotes the polarization vectors of photon and $Z$ boson. $u_t$ and $u_c$ denote the wave functions of the top and charm quarks, $p$ is the momentum of the top quark, $p'$ is the momentum of the charm quark and $q$ is the momentum of the vector boson. $m_{\tilde{D}_i}$, $m_{\tilde{D}_k}$ and $m_{\chi^\pm}$ are the mass eigenvalues from Eq.(\ref{Y2}). Correspondingly, $A_{\bar{c}\tilde{D}\chi^\pm}^L $,
$A_{\bar{c}\tilde{D}\chi^\pm}^R$,
$A_{\bar{\chi}^\pm\tilde{D}t}^L$ ,
$A_{\bar{\chi}^\pm\tilde{D}t}^R$ and
 $B_{V\tilde{D}\tilde{D}}$ are the coupling vertices \cite{32,33,39,40}. $L$ and $R$ in the subscripts denote the left-handed and right-handed parts, respectively. They can be derived from SARAH.

In the next calculation, we will first solve the Feynman integral,
and the formula we use for the integration of the denominator is \cite{66}:
\begin{eqnarray}
&&\frac{1}{ABC}=\int^1_0dx\int^1_02ydy\frac{1}{[(Ax+B(1-x))y+C(1-y)]^3}.\label{G8}
\end{eqnarray}

Calculating integral in this way can greatly increase the efficiency of numerical calculation in our work. According to Eq.(\ref{G8}) we get:
\begin{eqnarray}
&&\int^1_0dx\int^1_02ydy\Big\{\Big([(p\!\!\!\slash-k\!\!\!\slash)^2-m_{\chi^\pm_n}^2]x+(k^2-m_{\tilde{D}_i}^2)(1-x)\Big)y
+\Big((k\!\!\!\slash-q\!\!\!\slash)^2-m_{\tilde{D}_k}^2\Big)(1-y)\Big\}^{-3} \nonumber\\&&\hspace{0.2cm}
=\int^1_0dx\int^1_02ydy\Big\{[k-pxy-q(1-y)]^2-[pxy+q(1-y)]^2+(p^2-m_{\chi^\pm_n}^2)xy\nonumber\\&&\hspace{0.6cm}
-m_{\tilde{D}_i}^2(1-x)y+(q^2-m_{\tilde{D}_k}^2)(1-y)\Big\}^{-3},
 \end{eqnarray}
here we let $b\!\!\!\slash=-p\!\!\!\slash xy-q\!\!\!\slash(1-y)$, $k'\!\!\!\!\slash=k\!\!\!\slash+b\!\!\!\slash$ and $J=-[p^2(xy-x^2y^2)+q^2((1-y)-(1-y)^2)-2p\cdot qxy(1-y)-m_{\chi^\pm_n}^2xy-m_{\tilde{D}_i}^2(1-x)y-m_{\tilde{D}_k}^2(1-y)]$ to get the final form of the denominator in Eq.(\ref{K1}) as:
\begin{eqnarray}
&&\int^1_0dx\int^1_02ydy\frac{1}{(k^{'2}-J)^3}.
 \end{eqnarray}

This type of substitution is also performed for the numerator of Eq.(\ref{K1}). We take all the diagrams in Fig.\ref{A1} and calculate the Feynman amplitudes for each of $t\rightarrow c\gamma$, $t\rightarrow cg$ and $t\rightarrow cZ$.
Finally we perform the calculation of the respective mode squares for the three processes.

In the MSSM, the top quark decay $t\rightarrow ch$ is flavor-changing, where $h$ is the lightest CP-even Higgs boson. By studying Fig.\ref{A2}, we find that in addition to the new contribution of the down-type quarks, the mixing between the Higgs doublet and the exotic single-line states $\tilde{\eta}_{1,2}$ also affects the $t\rightarrow ch$ decay channel. We use the following calculation of Fig.\ref{A2}(4) as an example. The amplitude can be expressed as:
\begin{eqnarray}
&&\mathcal{M}_{t\rightarrow ch}=-\bar{u}_c(p')\int\frac{d^Dk}{(2\pi)^D}\frac{1}{[(k-q)^2-m_{\chi^\pm}^2](k^2-m_{\chi^\pm}^2)[(p-k)^2-m_{\tilde{D}_j}^2]}      ,\nonumber\\&&\hspace{1.7cm}
\cdot (A_{\bar{c}\chi^\pm\tilde{D}_j}^LP_L+A_{\bar{c}\chi^\pm\tilde{D}_j }^RP_R)(k\!\!\!\slash-q\!\!\!\slash+m_{\chi^\pm})(A_{\bar{\tilde{D}}_j^\dag\chi^\pm t}^LP_L+A_{\bar{\tilde{D}}_j^\dag\chi^\pm t}^RP_R)\nonumber\\&&\hspace{1.7cm}
\cdot(k\!\!\!\slash+m_{\chi^\pm})(A_{h\chi^\pm\chi^\pm }^LP_L+A_{h\chi^\pm\chi^\pm }^RP_R)u_t(p).
\end{eqnarray}
Other graphs of the $t\rightarrow ch$ process can be calculated similarly.

We use dimensional regularization to treat the divergences with $d=4-2\epsilon$ and the limit $d\rightarrow4$.
To obtain finite results, the divergences are canceled by the modified minimal substraction $(\overline{MS})$ scheme.
The terms proportional to $\frac{1}{\epsilon}-\gamma_E+\log(4\pi)$ are
deleted. Here $\gamma_E\approx0.5772$ is Euler constant.

Based on the above calculations, the branching ratios of the top quark rare decays are respectively:
\begin{eqnarray}
&&Br(t\rightarrow cV)=\frac{|\mathcal{M}_{tcV}|^2 \sqrt{((m_t+m_V)^2-m_c^2)((m+t-m_V)^2-m_c^2)}}{32\pi m_t^3\Gamma_{total}},\nonumber\\&&
Br(t\rightarrow ch)=\frac{|\mathcal{M}_{tch}|^2 \sqrt{((m_t+m_h)^2-m_c^2)((m+t-m_h)^2-m_c^2)}}{32\pi m_t^3\Gamma_{total}},
\end{eqnarray}
where $\Gamma_{total}$=$1.42_{-0.15}^{+0.19}$~{\rm GeV} \cite{5} is the total decay width of top quark.

\section{Numerical analysis}
In this section, we study the numerical results of flavor violation for the top-quark $t\rightarrow cV, ch$ processes. According to the latest LHC data \cite{43,44,45,46,47}, our values are subject to certain constraints. So we consider a number of individual experimental constraints including:

1. The lightest CP-even Higgs mass is around 125.25 GeV \cite{5,48,49,50}.

2. The updated experimental data show that the mass of the $Z'$ boson at the 95\% confidence level (CL) \cite{51} satisfies $M_{Z'}>$ 5.15 TeV.
Eq.(\ref{G8}) leads to an approximate result of $M_{Z'}$ as  $M_{Z'}\approx g_X\xi>$ 5.15 TeV.

3. The ratio between $M_{Z'}$ and its gauge coupling constant $\frac{M_{Z'}}{g_X} \geqslant $6~{\rm TeV} \cite{52,53},
so $g_X$ is restricted in the region 0$<g_X\leqslant$0.85.

4. The new angle $\beta_\eta$ is constrained by LHC as $\tan\beta_\eta<1.5$ \cite{57}.

5. The limitations for the particle masses accord to the PDG \cite{5} data, and the concrete contents are the following. The neutralino mass is limited to more than 116 GeV, the chargino mass is limited to more than 1000 GeV and the scalar quark mass is greater than 1300 GeV.

The relevant SM input parameters in the numerical program are selected as: $m_Z$=91.188~{\rm GeV}, $m_W$=80.385~{\rm GeV}, $m_c$=1.27~{\rm GeV}, $m_t$=172.69~{\rm GeV}. In conjunction with the above experimental requirements, we obtain a wealth of data and use graphs to analyze and process the data. We generally take the values of new particle masses($M_{BB'},~M_{BL}$) near the order of $10^3$ GeV, which is around the energy scale of new physics.
$T_{\lambda_C}$  and $T_{\lambda_H}$ are trilinear coupling coefficients, which are roughly in the order of magnitude of the mass, and can be varied up or down to the order of $10^2$ $\sim$ $10^4$ GeV. $M_{Uii}^2$, $M_{Qii}^2$, $M_{Dii}^2$, $B_S$ and $B_\mu$ are all of mass square dimension, and
can be up to the order of $10^6$ GeV$^2$. The dimensionless parameters $\lambda_C$ and $\lambda_H$ are generally taken as numbers less than 1. Considering the above constraints in the previous paragraphs, we use the following parameters:
\begin{eqnarray}
&&M_{BB'}=400~{\rm GeV},~M_{Uii}^2=6\times10^6~{\rm GeV^2} (i=1,2,3),~T_{\lambda_C}=-100~{\rm GeV},\nonumber\\&&
M_{BL}=1000~{\rm GeV},~\lambda_C=-0.08,~T_{\lambda_H}=300~{\rm GeV},~\kappa=0.1,\nonumber\\&&
l_W=4\times10^6~{\rm GeV^2},~B_\mu=B_S=1\times10^6~{\rm GeV^2}.
\end{eqnarray}

Here, we default to the non-diagonal elements of the mass matrix being set to zero if not otherwise specified. In $U(1)_X$SSM,~$g_{YX}$ is the mixing gauge coupling constant of $U(1)_Y$ group and $U(1)_X$ group, and it is the parameter beyond MSSM. The mass matrices of neutralino, down type squark and up type squark all contains $g_{YX}$. Furthermore, $g_{YX}$ appears in the vertex, which can enlarge the coupling constant of the vertex. $M_{BB'}$ is the mass of the $U(1)_Y$ and $U(1)_X$ gaugino mixing and presents in the mass matrix of neutralino. $\tan\beta$ appears in almost all the mass matrices of fermions, scalars and Majoranas, and it must be a sensitive parameter. It can affect the masses of particles and vertex couplings by directly affecting $v_u$ and $v_d$, $M_{BL}$ is the mass of the new gaugino, and it has influence on the mass matrix of neutralino. $\lambda_H$ relates to the strength of the self-interaction coupling of the Higgs field, which affects the VEV and the Higgs boson mass.

\subsection{The process of $t\rightarrow c\gamma $}
In order to find out the parameters affecting the top quark flavor violation, some sensitive parameters need to be studied. To show the numerical results clearly, the parameters are set to $M_{Dii}^2=6\times10^6~{\rm GeV^2}$, $M_{Qii}^2=6\times10^6~{\rm GeV^2}$ (i=1,2,3), $\mu=1000~{\rm GeV}$, $M_1=1200~{\rm GeV}$.
We plot the relationship between Br($t\rightarrow c\gamma $) and different parameters.

\begin{figure}[ht]
\setlength{\unitlength}{5mm}
\centering
\includegraphics[width=3in]{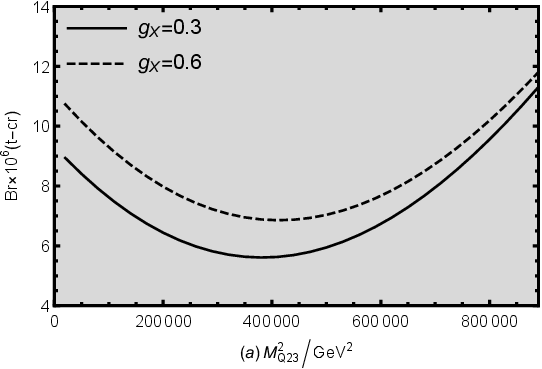}
\setlength{\unitlength}{5mm}
\centering
\includegraphics[width=3.07in]{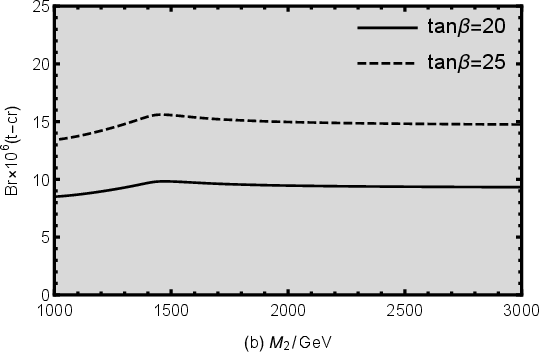}
\setlength{\unitlength}{5mm}
\centering
\includegraphics[width=2.96in]{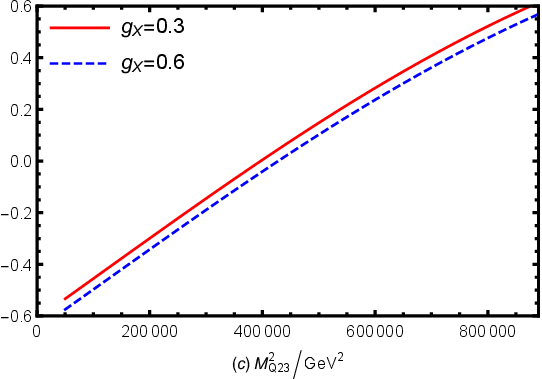}
\setlength{\unitlength}{5mm}
\centering
\includegraphics[width=3.05in]{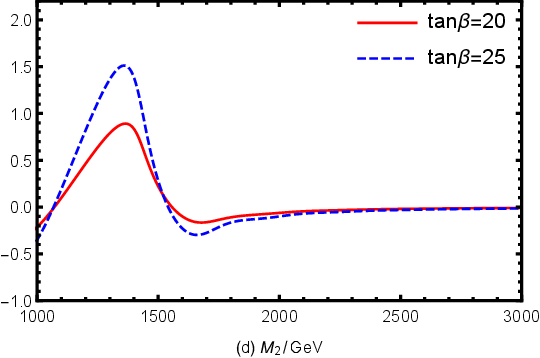}

\caption{The diagrams of Br($t\rightarrow c\gamma$) affected by different parameters. The gray area is reasonable value range, where Br($t\rightarrow c\gamma$) is lower than the upper limit. The solid line and dashed line in Fig.\ref{T1}(a) correspond to $g_{X}=0.3$ and $g_{X}=0.6$. The solid line and dashed line in Fig.\ref{T1}(b) correspond to $\tan\beta=20$ and $\tan\beta=25$, as $M_{Qij}=10^5 ~{\rm GeV^2}$ $(i=j=1,2,3, i\neq j)$. Fig.\ref{T1}(c) shows the differential distribution of (a) and Fig.\ref{T1}(d) shows the differential distribution of (b).}{\label {T1}}
\end{figure}

First we plot the one-dimensional diagrams of $Br(t\rightarrow c\gamma)$ versus $M_{Q23}^2$, $M_2$ as shown in Fig.\ref{T1}. The gray shaded area is the experimental limit satisfied by the $Br(t\rightarrow c\gamma)$ process. In Fig.\ref{T1} (a) we plot $Br(t\rightarrow c\gamma)$ versus $M_{Q23}^2$. Let $\tan\beta=20$, $M_2=1200~{\rm GeV}$, $g_{YX}=0.2$ and $\lambda_H=0.1$, the solid line corresponds to $g_{X}$=0.3 and the dashed line corresponds to $g_{X}$=0.6. Overall, both  lines show a decreasing trend in the range of $0-4\times10^5~{\rm GeV^2}$ for $M_{Q23}^2$ due to the fact that the contribution of the lower-type squarks is canceled by the contribution of the charged Higgs boson at the turning point. Then it is followed by an upward trend, i.e., which means that $Br(t\rightarrow c\gamma)$ increases as $M_{Q23}^2\geq4\times10^5~{\rm GeV^2}$. From bottom to top in the Fig.\ref{T1} (a) $Br(t\rightarrow c\gamma)$ increases as the value of $g_{X}$ increases. Fig.\ref{T1} (c) is the differential distribution of Fig.\ref{T1} (a). From Fig.\ref{T1} (c) we find that the trend and pattern of the values in Fig.\ref{T1} (a) are more obvious. In Fig.\ref{T1} (c), the differential increases linearly, and the speed of variable change is relatively smooth. It shows that $M_{Q23}^2$ is a parameter that has influence on $Br(t\rightarrow c\gamma)$.

In Fig.\ref{T1} (b) we plot $Br(t\rightarrow c\gamma)$ versus $M_2$, such that $M_{Qij}^2=10^5~{\rm GeV^2}$ $(i,j=1,2,3, i\neq j)$, $g_X=0.3$, $\lambda_H=0.1$, $g_{YX}=0.2$, the solid line corresponds to $\tan\beta=20$, and the dashed line corresponds to $\tan\beta=25$. There is a slight bulge in the $Br(t\rightarrow c\gamma)$ value at $M_2=1400~{\rm GeV}$, followed by a slight downward trend. It can be seen that the overall value satisfies this limit and its overall trend is a decreasing function. As the line in the graph goes from bottom to top i.e. as $\tan\beta$ increases $Br(t\rightarrow c\gamma)$ also increases gradually. Fig.\ref{T1} (d) is the differential distribution of Fig.\ref{T1} (b). In Fig.\ref{T1} (d) there is a maximum value of differential value appears when $M_2$=1400 GeV, at this time $Br(t\rightarrow c\gamma)$ is the maximum value. The differential value is negative when $M_2>$1600 GeV. That is to say, $Br(t\rightarrow c\gamma)$ is decreasing, but the decrease trend is very small.

In order to better and more deeply explore the parameter space, as $M_2=1200~{\rm GeV}$, we scan some parameters randomly, and in the $Br(t\rightarrow c\gamma)$ process we scan the following parameters:
\begin{eqnarray}
&&5\leq\tan\beta\leq50,~~0.3\leq g_X\leq0.7,~~0.01\leq g_{YX}\leq0.5,~~0.1\leq \lambda_H\leq0.4.
\end{eqnarray}

\begin{figure}[ht]
\setlength{\unitlength}{5mm}
\centering
\includegraphics[width=3in]{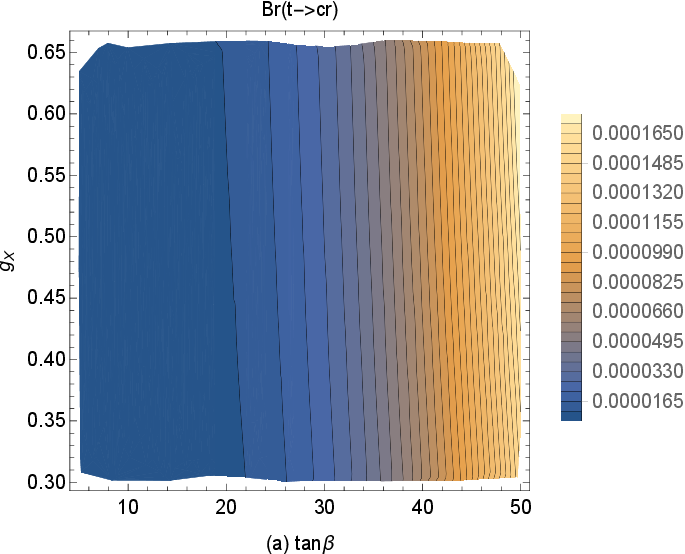}
\setlength{\unitlength}{5mm}
\centering
\includegraphics[width=3in]{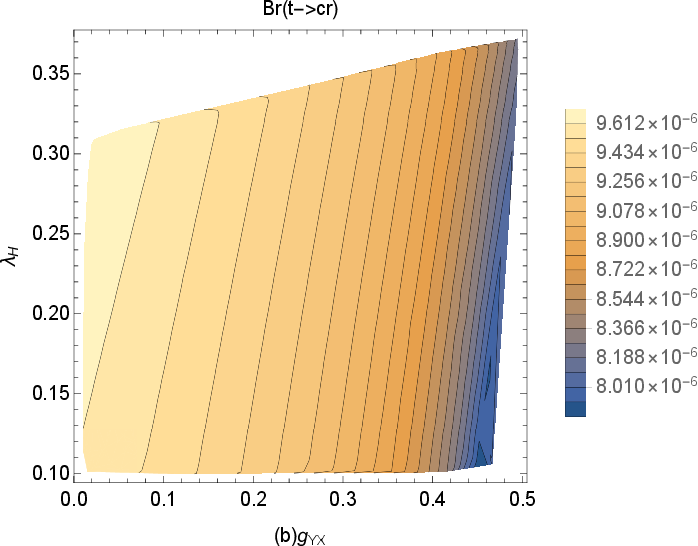}
\caption{(a) Effects of $\tan\beta$ and $g_X$ on $Br(t\rightarrow c\gamma)$. The horizontal coordinate indicates the range $5\leq\tan\beta\leq50$ and the vertical coordinate indicates $0.3\leq g_X\leq0.7$. (b) Effects of $g_{YX}$ and $\lambda_H$ on $Br(t\rightarrow c\gamma)$. The horizontal coordinate indicates the range $0.01\leq g_{YX}\leq0.5$ and the vertical coordinate indicates $0.1\leq \lambda_H\leq0.4$. The icons on the right side indicate the colors corresponding to the values of $Br(t\rightarrow c\gamma)$.}{\label {T2}}
\end{figure}

In Fig.\ref{T2}(a) we set $\lambda_H=0.1$, $g_{YX}=0.2$ to explore the effects of $\tan\beta$ and $g_{X}$ on $Br(t\rightarrow c\gamma)$. It is clear from the figure that the value of $Br(t\rightarrow c\gamma)$ increases as $\tan\beta$ increases. When $\tan\beta$ reaches its maximum value of 50, $Br(t\rightarrow c\gamma)$ reaches an order of magnitude of $10^{-4}$, very close to the experimental upper limit.
It indicates that $\tan\beta$ is a very important parameter. The value
of $Br(t\rightarrow c\gamma)$ turns large as $g_{X}$ increases, but the effect of $g_{X}$ on $Br(t\rightarrow c\gamma)$ is small and hardly noticeable.

In Fig.\ref{T2}(b) we set $\tan\beta=20$ and $g_{X}=0.3$ to explore the effects of $\lambda_H$ and $g_{YX}$ on $Br(t\rightarrow c\gamma)$. From the Fig.\ref{T2}(b) we find that both $\lambda_H$ and $g_{YX}$ have effects on $Br(t\rightarrow c\gamma)$. The value of $Br(t\rightarrow c\gamma)$ decreases with the increase of $g_{YX}$, and the smaller the value of $g_{YX}$ the closer to the upper limit of the $Br(t\rightarrow c\gamma)$. There is some slight increase in the value of $Br(t\rightarrow c\gamma)$ with the enlaring $\lambda_H$, but the impact of $\lambda_H$ is relatively small compared to $g_{YX}$. In Fig.\ref{T2}(b), we find the presence of a white area in the upper left corner. This occurs because of the limitation from the masses of Higgs and other particles.

\subsection{The process of $t\rightarrow cg$}
\begin{figure}[ht]
\setlength{\unitlength}{5mm}
\centering
\includegraphics[width=3in]{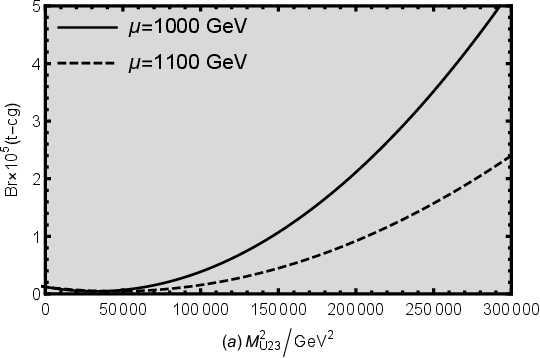}
\setlength{\unitlength}{5mm}
\centering
\includegraphics[width=2.95in]{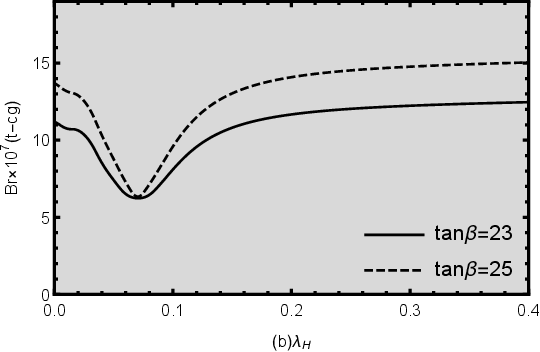}
\setlength{\unitlength}{5mm}
\centering
\includegraphics[width=2.9in]{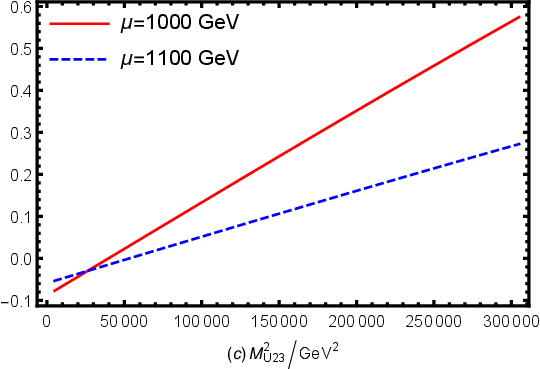}
\setlength{\unitlength}{5mm}
\centering
\includegraphics[width=2.9in]{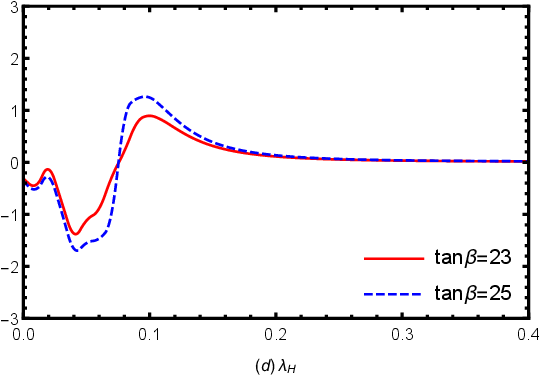}

\caption{Br($t\rightarrow cg$) diagrams affected by different parameters. The gray area is reasonable value range, where Br($t\rightarrow cg$) is lower than the upper limit. The solid line and dashed line in Fig.\ref{T3}(a) correspond to $\mu=1000~{\rm GeV}$ and $\mu=1100~{\rm GeV}$. The solid line and dashed line in Fig.\ref{T3}(b) correspond to $\tan\beta=23$ and $\tan\beta=25$. Fig.\ref{T3}(c) shows the differential distribution of (a) and Fig.\ref{T3}(d) shows the differential distribution of (b).}{\label {T3}}
\end{figure}

In this section we continue our exploration of the branching ratio of $t\rightarrow cg$ with respect to certain parameters, we make $M_1=1200~{\rm GeV}$, $M_2=1200~{\rm GeV}$, $M_{Dii}^2=6\times10^6~{\rm GeV^2}$, $M_{Qii}^2=6\times10^6~{\rm GeV^2}$. Here we plot the one-dimensional diagrams of Br($t\rightarrow cg$) in Fig.\ref{T3}. The gray shaded portion continues to indicate the experimental limit that is satisfied by the Br($t\rightarrow cg$) process. In Fig.\ref{T3} (a) we plot Br($t\rightarrow cg$) versus $M_{U23}^2$, as $\tan\beta=20$, $\lambda_H$=0.1, $g_{X}=0.3$, $g_{YX}=0.2$,
where the solid and dashed lines correspond to $\mu=1000~{\rm GeV}$ and $\mu=1100~{\rm GeV}$, respectively. We find that Br($t\rightarrow cg$) increases with $M_{U23}^2$ and the value approaches the experimental upper limit more as $M_{U23}^2$
 further increases. We also find that Br($t\rightarrow cg$) decreases as $\mu$ increases. Fig.\ref{T3} (c) shows the differential distribution of Fig.\ref{T3} (a). From Fig.\ref{T3} (c) we find that the variation of Fig.\ref{T3} (a) is quite regular. We clearly see that $t\rightarrow cg$ is increasing with the increase of $M_{U23}^2$.

In Fig.\ref{T3} (b) we plot Br($t\rightarrow cg$) versus $\lambda_H$ with $M_{Q23}=10^5~{\rm GeV^2}$, $\mu=1000~{\rm GeV}$, $g_{X}=0.3$ and $g_{YX}=0.2$. With $\tan\beta=23$ (solid line) and $\tan\beta=25$ (dashed line). From the plot we find that the Br($t\rightarrow cg$) value shows a minimum at $\lambda_H=0.08$, which is due to the mixing of several parameters. Its overall value satisfies the experimental limit of the process and is five to six orders of magnitude higher than the SM prediction. As $\tan\beta$ increases from 23 to 25, the Br($t\rightarrow cg$) value also increases, but the Br($t\rightarrow cg$) values almost coincide at $\lambda_H=0.07$. Fig.\ref{T3} (d) is the differential distribution of Fig.\ref{T3} (b). In Fig.\ref{T3} (d) the differential values are negative when $\lambda_H<$0.08, at this time the Br($t\rightarrow cg$) is decreasing as $\lambda_H$ becomes larger. In the range of $0.08\leq\lambda_H<0.11$, the slope is the largest, at this time the Br($t\rightarrow cg$) changes faster than the others. When $\lambda_H\geq$0.11, the differential values are all positive, but the value change is small, at this time the Br($t\rightarrow cg$) is still getting bigger, only the magnitude of the increase is smaller.

Let $\mu=1000~{\rm GeV}$ and $\lambda_H=0.1$. Next we will randomly scan the parameters $\tan\beta$, $g_X$, $g_{YX}$ and the ranges are:
\begin{eqnarray}
&&5\leq\tan\beta\leq50,~~0.3\leq g_X\leq0.7,~~0.01\leq g_{YX}\leq0.5.
\end{eqnarray}

\begin{figure}[ht]
\setlength{\unitlength}{5mm}
\centering
\includegraphics[width=3in]{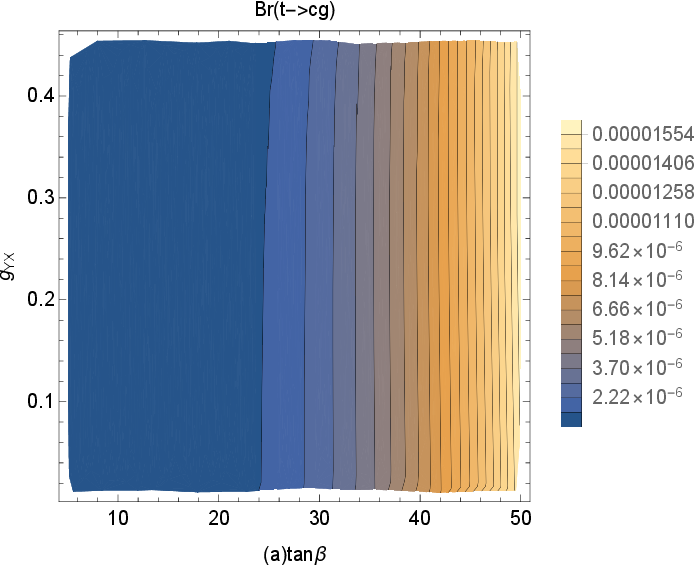}
\setlength{\unitlength}{5mm}
\centering
\includegraphics[width=3in]{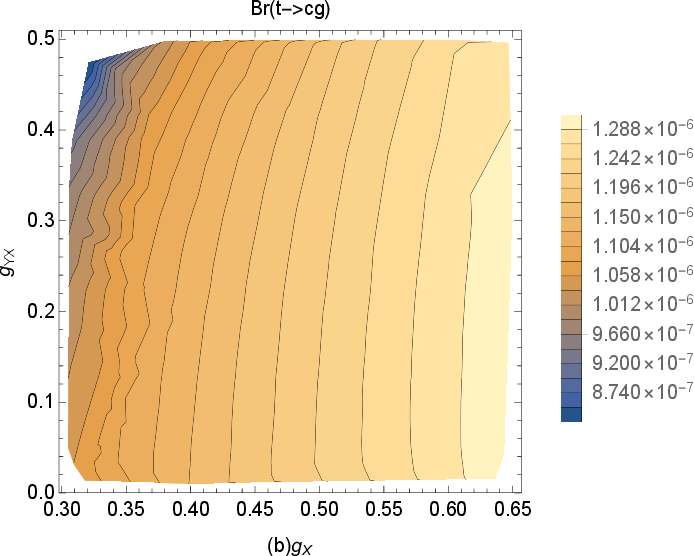}
\caption{(a) Effects of $\tan\beta$ and $g_{YX}$ on $Br(t\rightarrow cg)$. The horizontal coordinate indicates the range $5\leq\tan\beta\leq50$ and the vertical coordinate indicates $0.01\leq g_{YX}\leq0.5$. (b) Effects of $g_{X}$ and $g_{YX}$ on $Br(t\rightarrow cg)$. The horizontal coordinate indicates the range $0.3\leq g_{X}\leq0.7$ and the vertical coordinate indicates $0.01\leq g_{YX}\leq0.5$. The icons on the right side indicate the colors corresponding to the values of $Br(t\rightarrow cg)$.}{\label {T4}}
\end{figure}
In Fig.\ref{T4} (a), we set $g_X=0.3$ to explore the effects of $\tan\beta$ and $g_{YX}$ on $Br(t\rightarrow cg)$. We find that as $\tan\beta$ increases $Br(t\rightarrow cg)$ gradually changes from blue to yellow, i.e., a significant increase in $Br(t\rightarrow cg)$ occurs, and the larger the $\tan\beta$ the closer $Br(t\rightarrow cg)$ gets to the experimental upper limit.
 $g_{YX}$  has some minor effect on the results, and it is not apparent enough here. In Fig.\ref{T4} (b), we set $\tan\beta=20$ to explore the effects of $g_{X}$ and $g_{YX}$ on $Br(t\rightarrow cg)$, and it shows that the larger $g_{X}$ is the larger $Br(t\rightarrow cg)$ is. On the other hand the larger $g_{YX}$ is the smaller $Br(t\rightarrow cg)$ is. The $Br(t\rightarrow cg)$ is maximized when $g_{X}=0.7$ and $g_{YX}=0.01$, and the $Br(t\rightarrow cg)$ is closer to the upper limit of the experiment. When $g_{YX}$ tends to zero, the dependence of the branching ratio on $g_{X}$ is strong.

\subsection{The process of $t\rightarrow cZ$}
\begin{figure}[ht]
\setlength{\unitlength}{5mm}
\centering
\includegraphics[width=3.05in]{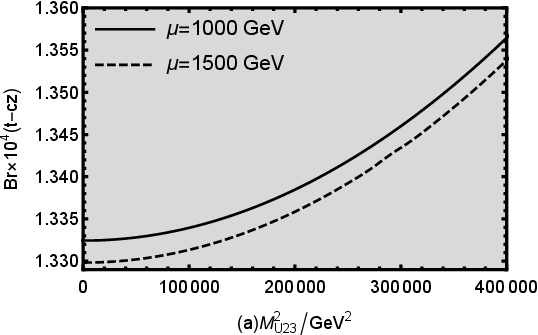}
\setlength{\unitlength}{5mm}
\centering
\includegraphics[width=2.9in]{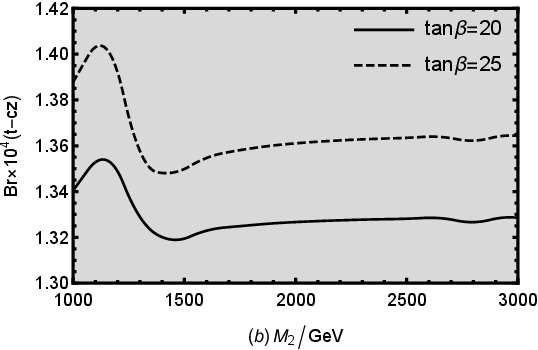}
\setlength{\unitlength}{5mm}
\centering
\includegraphics[width=2.95in]{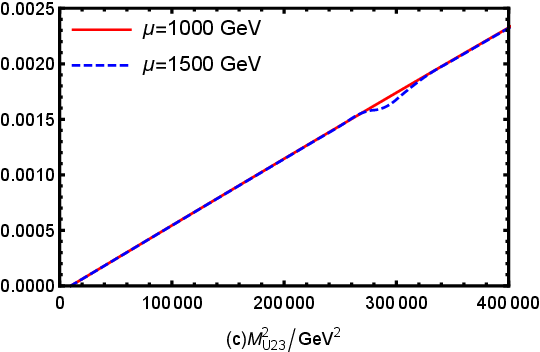}
\setlength{\unitlength}{5mm}
\centering
\includegraphics[width=2.9in]{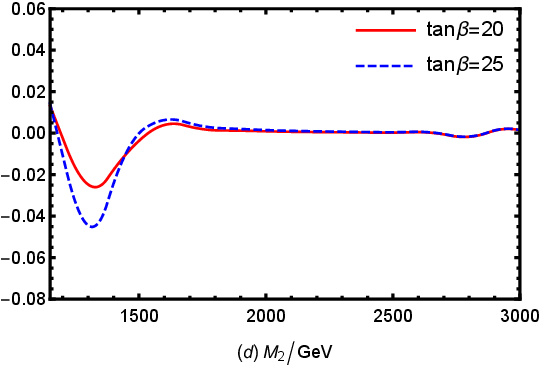}
\caption{Br($t\rightarrow cZ $) diagrams affected by different parameters. The gray area is reasonable value range, where $Br(t\rightarrow cZ)$ is lower than the upper limit. The solid line and dashed line in Fig.\ref{T5}(a) correspond to $\mu=1000~{\rm GeV}$ and $\mu=1500~{\rm GeV}$. The solid line and dashed line in Fig.\ref{T5}(b) correspond to $\tan\beta=20$ and $\tan\beta=25$. Fig.\ref{T5}(c) shows the differential distribution of (a) and Fig.\ref{T5}(d) shows the differential distribution of (b).}{\label {T5}}
\end{figure}

\begin{figure}[ht]
\setlength{\unitlength}{5mm}
\centering
\includegraphics[width=3in]{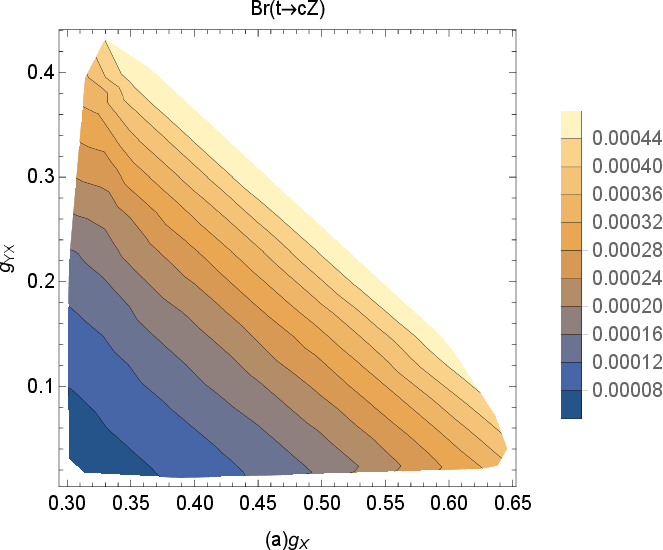}
\setlength{\unitlength}{5mm}
\centering
\includegraphics[width=3in]{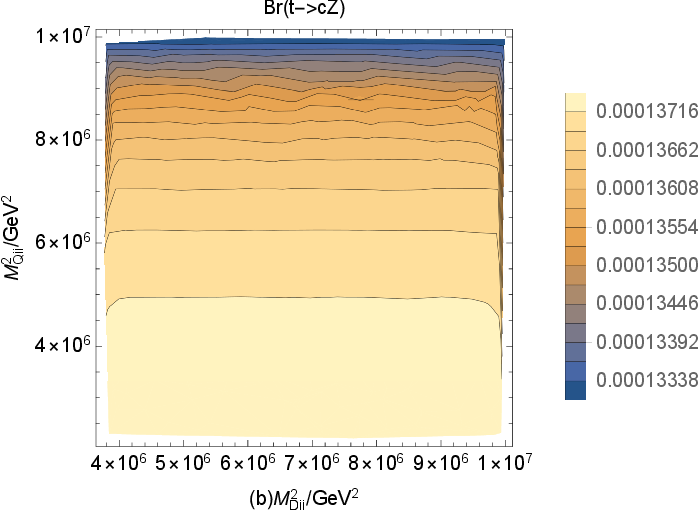}
\caption{(a) Effects of $g_{X}$ and $g_{YX}$ on $Br(t\rightarrow cZ)$. The horizontal coordinate indicates the range $0.3\leq g_{X}\leq0.7$ and the vertical coordinate indicates $0.01\leq g_{YX}\leq0.5$. (b) Effects of $M_{Dii}^2$ and $M_{Qii}^2$ on $Br(t\rightarrow cZ)$. The horizontal coordinate indicates the range $10^6~{\rm GeV^2}\leq M_{Dii}^2\leq10^7~{\rm GeV^2}$ and the vertical coordinate indicates $10^6~{\rm GeV^2}\leq M_{Qii}^2\leq10^7~{\rm GeV^2}$. The icons on the right side indicate the colors corresponding to the values of $Br(t\rightarrow cZ)$.}{\label {T6}}
\end{figure}

The experimental upper bound ($5\times10^{-4}$) for the $Br(t\rightarrow cZ)$ process is of the same order of magnitude as for $Br(t\rightarrow c\gamma)$ and $Br(t\rightarrow cg)$. In this subsection, we still discuss the effects of different parameters on the $Br(t\rightarrow cZ)$ branching ratio, where we focus on the effects of the parameters $M_{U23}^2$, $M_2$, $g_X$, $g_{YX}$, $M_{Dii}^2$ and $M_{Qii}^2$. We fix the value of the parameters $M_1=1200~{\rm GeV}$ and $\lambda_H=0.1$, one-dimensional Fig.\ref{T5} and multidimensional plots Fig.\ref{T6} are drawn.

In Fig.\ref{T5} (a) we plot $Br(t\rightarrow cZ)$ versus $M_{U23}^2$ and make $M_2=1200~{\rm GeV}$, $\tan\beta=20$, $g_X=0.3$, $g_{YX}=0.2$, $M_{Dii}^2=6\times10^6~{\rm GeV^2}$, $M_{Qii}^2=6\times10^6~{\rm GeV^2}$, with the solid and dashed lines representing $\mu=1000~{\rm GeV}$ and $\mu=1500~{\rm GeV}$, respectively. It is easy to notice that both lines show an increasing trend, i.e., $Br(t\rightarrow cZ)$ increases with increasing $M_{U23}^2$, but the trend of change is not very large in terms of value. Also we find a decrease in $Br(t\rightarrow cZ)$ as $\mu$ increases. Both curves are shaded in gray. In Fig.\ref{T5} (b) we fix $\mu=1000~{\rm GeV}$, $g_X=0.3$, $g_{YX}=0.2$, $M_{Dii}^2=6\times10^6~{\rm GeV^2}$, $M_{Qii}^2=6\times10^6~{\rm GeV^2}$ and plot $Br(t\rightarrow cZ)$ versus $M_2$, with $\tan\beta=20$ (solid line) and $\tan\beta=25$ (dashed line). The two lines are convex and reach a maximum at $M_2=1150~{\rm GeV}$, then show a downward trend and finally level off, i.e., in the range of $1000~{\rm GeV}\leq M_2\leq1500~{\rm GeV}$, $M_2$ has a obvious influence on $Br(t\rightarrow cZ)$. The solid and dashed lines run from bottom to top, indicating that $\tan\beta$ is also a sensitive parameter for $Br(t\rightarrow cZ)$, which increases with $\tan\beta$.

Fig.\ref{T5} (c) shows the differential distribution of Fig.\ref{T5} (a). In Fig.\ref{T5} (c) the difference values are all positive. When $\mu=1000~{\rm GeV}$ and $\mu=1500~{\rm GeV}$, the difference between the values is very small, so the two lines almost coincide. Fig.\ref{T5} (d) shows the differential distribution of Fig.\ref{T5} (b). In Fig.\ref{T5} (d), the value of differential in most regions is negative when $M_2 < 1500~{\rm GeV}$, and all other values are positive.

In order to explore the effects of different parameters on $Br(t\rightarrow cZ)$ in greater depth, let $\mu=1000~{\rm GeV}$, $M_2=1200~{\rm GeV}$, $\tan\beta=20$, we will continue to perform randomized scans with different parameters in the range:
\begin{eqnarray}
&&0.01\leq g_{YX}\leq0.5,~~10^6~{\rm GeV^2}\leq M_{Dii}^2\leq10^7~{\rm GeV^2},\nonumber\\&&0.3\leq g_X\leq0.7,~~10^6~{\rm GeV^2}\leq M_{Qii}^2\leq10^7~{\rm GeV^2}~(i=1,2,3).
\end{eqnarray}

In Fig.\ref{T6} (a) we set $M_{Qii}^2=6\times10^6~{\rm GeV^2}$, $M_{Dii}^2=6\times10^6~{\rm GeV^2}$ to explore the effects of $g_X$ and $g_{YX}$ on $Br(t\rightarrow cZ)$. We can clearly see that $Br(t\rightarrow cZ)$ has a minimum at $g_X=0.3$ and $g_{YX}=0.01$. For $Br(t\rightarrow cZ)$ both $g_X$ and $g_{YX}$ are sensitive parameters. $g_{YX}$ is a coupling constant that affects the strength of gauge mixing. Further more $g_X$ and $g_{YX}$ make a new contribution to $Br(t\rightarrow cZ)$ through $Z-Z'$ mixing. $g_X$ also has a large effect on $Br(t\rightarrow cZ)$ when $g_{YX}$ tends to a minimum. $Br(t\rightarrow cZ)$ increases with enlarging $g_X$ and $g_{YX}$ vice versa. In Fig.\ref{T6} (a), the white region appears in the upper right corner. By analysing, we get that it is excluded by the experimental upper limit $Br(t\rightarrow cZ)<5\times10^{-4}$.

In Fig.\ref{T6} (b) we set $g_X=0.3$ and $g_{YX}=0.2$ to explore the effects of $M_{Qii}^2$ and $M_{Dii}^2$ on $Br(t\rightarrow cZ)$. The values of $M_{Qii}^2$ and $M_{Dii}^2$ are both in the range of $10^6~{\rm GeV^2}-10^7~{\rm GeV^2}$, and through Fig.\ref{T6} (b)
we find that $M_{Dii}^2$ has tiny effect on $Br(t\rightarrow cZ)$. However, as $M_{Qii}^2$ increases, $Br(t\rightarrow cZ)$ changes from yellow to blue, i.e., $Br(t\rightarrow cZ)$ decreases weakly as $M_{Qii}^2$ increases.

In summary, for the $t\rightarrow cZ$ process, the main sensitive parameters are  $M_2$, $g_X$, $g_{YX}$ and the off-diagonal element $M_{U23}^2$. While the diagonal elements $M_{Qii}^2$ (i=1,2,3) have an effect but not very sensitive.

\subsection{The process of $t\rightarrow ch$}
\begin{figure}[ht]
\setlength{\unitlength}{5mm}
\centering
\includegraphics[width=3in]{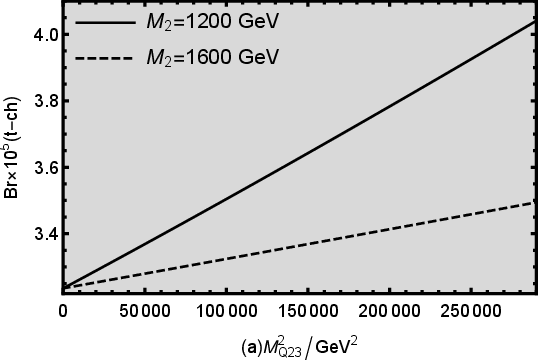}
\setlength{\unitlength}{5mm}
\centering
\includegraphics[width=3in]{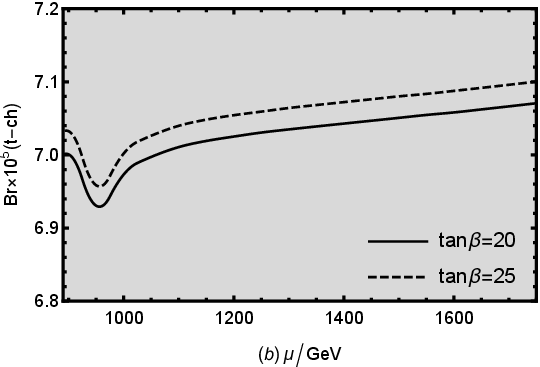}
\setlength{\unitlength}{4.9mm}
\centering
\includegraphics[width=2.9in]{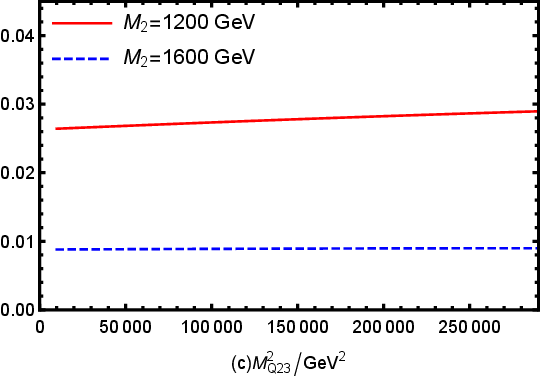}
\setlength{\unitlength}{5mm}
\centering
\includegraphics[width=3in]{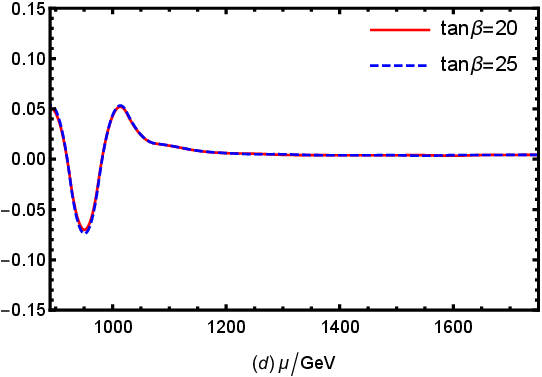}
\caption{Br($t\rightarrow ch$) diagrams affected by different parameters. The gray area is reasonable value range, where $Br(t\rightarrow ch)$ is lower than the upper limit. As $g_X=0.4$, The solid line and dashed line in Fig.\ref{T7}(a) correspond to $M_1=1200~{\rm GeV}$ and $M_1=1600~{\rm GeV}$. Making $M_{Q23}^2=10^5~{\rm GeV^2}$, the solid line and dashed line in Fig.\ref{T7}(b) correspond to $\tan\beta=20$ and $\tan\beta=25$. Fig.\ref{T7}(c) shows the differential distribution of (a) and Fig.\ref{T7}(d) shows the differential distribution of (b).}{\label {T7}}
\end{figure}

The experimental upper bound for $Br(t\rightarrow ch)$ is $1.1\times10^{-3}$, and in the $U(1)_X$SSM $Br(t\rightarrow ch)$ can be as high as $10^{-4}$ in some parts of the parameter space. Here we make $M_2=1200~{\rm GeV}$, $M_{Qii}^2=6\times10^6~{\rm GeV^2}$, $M_{Dii}^2=6\times10^6~{\rm GeV^2}$. We study the effects of parameters $M_{Q23}^2$, $\mu$, $M_1$ and $\tan\beta$ on $Br(t\rightarrow ch)$ in Fig.\ref{T7}. In Fig.\ref{T7} (a) we plot $Br(t\rightarrow ch)$ versus $M_{Q23}^2$ with $g_X=0.4$, $\mu=1000~{\rm GeV}$, $\tan\beta=20$, $g_{YX}=0.2$, $\lambda_H=0.1$. The solid line represents $M_1=1200~{\rm GeV}$ and the dashed line represents $M_1=1600~{\rm GeV}$. As the dashed and solid lines go from bottom to top, we find that $M_2$ has an effect on $Br(t\rightarrow ch)$ as sensitive parameter, and $Br(t\rightarrow ch)$ decreases as $M_2$ increases.
We can clearly find $Br(t\rightarrow ch)$ increases with the increase of $M_{Q23}^2$ and the trend of increase is almost linear. So $M_{Q23}^2$ is also a parameter that has strong effect on $Br(t\rightarrow ch)$. Fig.\ref{T7} (c) shows the differential distribution of Fig.\ref{T7} (a). In Fig.\ref{T7} (c), the differential values are all positive. The trend of the values is relatively smooth and the regularity is more obvious.

In Fig.\ref{T7} (b) we plot $Br(t\rightarrow ch)$ versus $\mu$ and fix $M_1=1200~{\rm GeV}$, $g_{X}=0.3$,  $g_{YX}=0.2$, $\lambda_H=0.1$, with the solid line representing $\tan\beta=20$ and the dashed line representing $\tan\beta=25$. By analyzing the previous three processes $t\rightarrow c\gamma$, $t\rightarrow cg$ and $t\rightarrow cZ$, we have concluded that $\tan\beta$ is one of the sensitive parameters. In the $t\rightarrow ch$ process we find from Fig.\ref{T7} (b) that $\tan\beta$ is still a more sensitive parameter. When $\tan\beta$ increases from 20 to 25, $Br(t\rightarrow ch)$ also increases as $\tan\beta$ increases.
It implies that $Br(t\rightarrow ch)$  shows a minimum at $\mu=950~{\rm GeV}$. When $\mu>950~{\rm GeV}$, $Br(t\rightarrow ch)$ increases as $\mu$ increases, but the trend of increase is relatively small. Fig.\ref{T7} (d) shows the differential distribution of Fig.\ref{T7} (b). In Fig.\ref{T7} (d), the trend of differential values is decreasing and then increasing when $\mu<1050~{\rm GeV}$. When $\mu\geq1050~{\rm GeV}$, the tendency of these two lines to decrease is getting weaker and weaker.

In order to further explore the effect of some other parameters on $Br(t\rightarrow ch)$, we make $M_1=1200~{\rm GeV}$, $\mu=1000~{\rm GeV}$ and perform a randomized scan of $g_X$, $g_{YX}$, $\lambda_H$, $\tan\beta$ as follows:
\begin{eqnarray}
&&0.01\leq g_{YX}\leq0.5,~~0.3\leq g_X\leq0.7,~~0.1\leq\lambda_H\leq0.4,~~10\leq\tan\beta\leq50.\label{S1}
\end{eqnarray}

\begin{figure}[h]
\setlength{\unitlength}{5mm}
\centering
\includegraphics[width=2.9in]{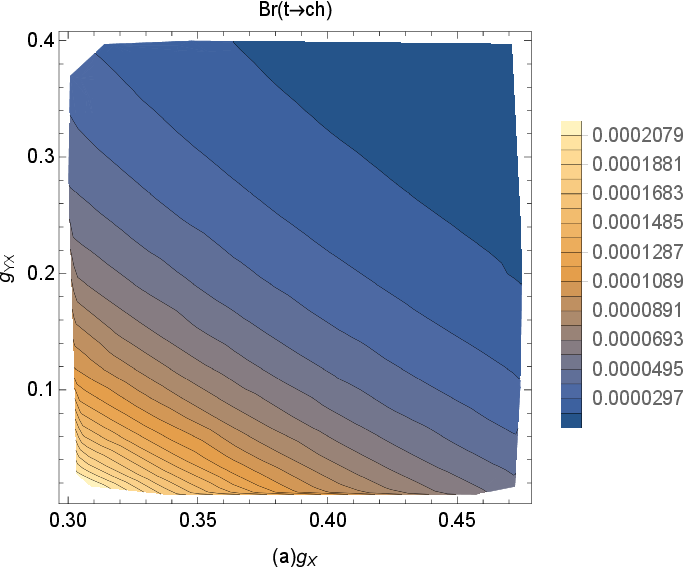}
\setlength{\unitlength}{5mm}
\centering
\includegraphics[width=2.9in]{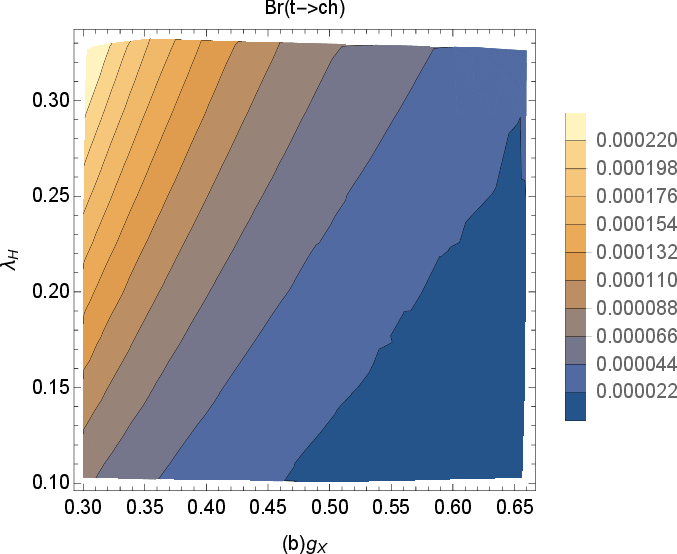}
\setlength{\unitlength}{5mm}
\centering
\includegraphics[width=2.9in]{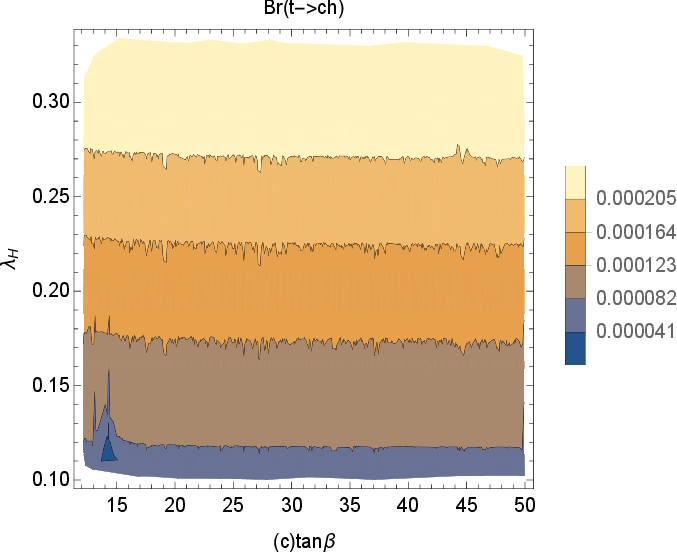}
\caption{(a) Effects of $g_{X}$ and $g_{YX}$ on $Br(t\rightarrow ch)$. The horizontal coordinate indicates the range $0.3\leq g_{X}\leq0.7$ and the vertical coordinate indicates $0.01\leq g_{YX}\leq0.5$. (b) Effects of $g_X$ and $\lambda_H$ on $Br(t\rightarrow ch)$. The horizontal coordinate indicates the range $0.3\leq g_X\leq0.7$ and the vertical coordinate indicates $0.1\leq\lambda_H\leq0.4$. (c) Effects of $\tan\beta$ and $\lambda_H$ on $Br(t\rightarrow ch)$. The horizontal coordinate indicates the range $5\leq \tan\beta\leq50$ and the vertical coordinate indicates $0.1\leq\lambda_H\leq0.4$. The icons on the right side indicate the colors corresponding to the values of $Br(t\rightarrow cZ)$.}{\label {T8}}
\end{figure}

Based on the parameters in Eq.(\ref{S1}), we obtain the data and plot Fig.\ref{T8}. In Fig.\ref{T8} (a) we explore the effect of $g_{X}$ ($0.3\leq g_{X}\leq0.7$) and $g_{YX}$ ($0.01\leq g_{YX}\leq0.5$) on $Br(t\rightarrow ch)$ by setting $\tan\beta=20$, $\lambda_H=0.1$. In Fig.\ref{T8} (b) we explore the effect of $g_{X}$ ($0.3\leq g_{X}\leq0.7$) and $\lambda_H~(0.1\leq\lambda_H\leq0.4)$ on $Br(t\rightarrow ch)$ by setting $\tan\beta=20$, $g_{YX}=0.2$. In Fig.\ref{T8} (c) we again set $g_{X}=0.3$ and $g_{YX}=0.2$ to explore the effect of $\tan\beta~(10\leq\tan\beta\leq50)$ and $\lambda_H~(0.1\leq\lambda_H\leq0.4)$ on $Br(t\rightarrow ch)$. Combining the three plots in Fig.\ref{T8}, we can clearly see that all four parameters $g_X$, $g_{YX}$, $\lambda_H$ and $\tan\beta$ affect the $Br(t\rightarrow ch)$, but they do so in different ways.
$g_{X}$ and $g_{YX}$ both present obvious effects on the numerical results.
$Br(t\rightarrow ch)$ is decreasing function of $g_{X}$ and $g_{YX}$. At the time they both take the minimum value, the branching ratio of the $t\rightarrow ch$ process reaches $10^{-4}$ which is very close to the experimental upper limit. In Fig.\ref{T8} (b) and Fig.\ref{T8} (c) it can be seen that $\lambda_H$ also behaves very sensitively, with $Br(t\rightarrow ch)$ increasing as $\lambda_H$ increases. In Fig.\ref{T8} (c), When $\tan\beta>10$, $\tan\beta$ has weak effect. This occurs because $\tan\beta$ not only appears in the diagonal sectors of the mass matrix, but also dominates the non-diagonal sectors, leading to the above results.

\section{Conclusion}
To summarize, in this work we study the rare decays $t\rightarrow c\gamma,~cg,~cZ,~ch$ of the top quark in the $U (1)_X$SSM. Compared to the MSSM, in the $U(1)_X$SSM we add three new Higgs singlets $\hat{\eta},~\hat{\bar{\eta}},~\hat{S}$ and three generations of right-handed neutrinos $\hat{\nu}_i$. Its local gauge group is $SU(3)_C\times SU(2)_L \times U(1)_Y\times U(1)_X$. Probing with the $U(1)_X$SSM makes our study richer and more interesting and provides a firmer basis for the existence of new physics. We start with one-loop diagrams and compute the Feynman amplitude for each process, to give numerical results for $Br(t\rightarrow c\gamma),~Br(t\rightarrow cg),~Br(t\rightarrow cZ),~Br(t\rightarrow ch)$. We try the effect of various parameters on the branching ratios of each process, select some clearer results, and plot one-dimensional and multidimensional plots for a more comprehensive numerical analysis based on experimental constraints.

In the used parameter space, numerical results show that all these processes are very close to the experimental upper limit, reaching even the same order of magnitude as the experimental upper limit at good parameter values, and may be detected in future high-energy colliders. By analyzing the numerical results in the adopted parameter space, we can conclude that $\tan\beta$, $g_X$, $g_{YX}$, $\mu$, $M_2$, $\lambda_H$ and the off-diagonal parameters $M_{U23}^2$, $M_{Q23}^2$ are the sensitive parameters that have a large influence on the branching ratios of the $t\rightarrow c\gamma,~cg,~cZ,~ch$ processes. Among them, the influence of $\tan\beta$ is the most pronounced, being the main parameter of the $t\rightarrow c\gamma,~cg,~cZ,~ch$ processes, and $\tan\beta$ appears in the mass matrixes of almost all SUSY particles.
It is a fact that $\tan\beta$ affects the numerical results mainly by influencing the particle masses,
the corresponding rotational matrices and the coupling vertexes.
$g_{YX}$ is the coupling constant in gauge mixing and it is parameterized outside of the MSSM. In couplings where squarks appear, $g_X$ and $g_{YX}$ will affect the squark masses and the corresponding rotation matrices, resulting in an effect on $Br(t\rightarrow c\gamma,~cg,~cZ,~ch)$. In addition, for the $t\rightarrow cZ$ process, they generate some new contributions via $Z-Z'$ mixing.
Comparing with the data in Table \ref {B6}, we can see that the maximum value of $Br(t\rightarrow cV, ch)$ obtained in $U(1)_X$SSM is greater than that achieved in B$-$LSSM. Especially for $Br(t\rightarrow ch)$ it is even four orders of magnitude higher than the B$-$LSSM result. The $Br(t\rightarrow cV, ch)$  obtained in $U(1)_X$SSM is already very close to the experimental upper limit, which makes our study more fulfilling and meaningful.

\begin{table*}
\caption{ The Upper Limit($95\%$C.L.), $B-L$SSM and $U(1)_X$SSM bounds on the decays $t\rightarrow cV, ch$}
\begin{tabular*}{\textwidth}{@{\extracolsep{\fill}}|l|l|l|l|l|@{}}
\hline
Decay&$t\rightarrow c\gamma$&$t\rightarrow cg$&$t\rightarrow cZ$&$t\rightarrow ch$\\
\hline
Upper Limit($95\%$C.L.)&$1.8\times10^{-4}$~~~~~~~&$2\times10^{-4}$~~~~~~~&~$5\times10^{-4}$~~~~~~~&$~1.1\times10^{-3}$~~~~~~~\\
\hline
$B-L$SSM \cite{21}&$5\times10^{-7}$&$2\times10^{-6}$&~$4\times10^{-7}$~&~$3\times10^{-9}~~~~~~$\\
\hline
$U(1)_X$SSM&$1.5\times10^{-4}$&$5\times10^{-5}$&~$1.41\times10^{-4}$&~$7.04\times10^{-5}$~~~~~~\\
\hline
\end{tabular*}
\label{B6}
\end{table*}

In conclusion, we have calculated the rare top-quark decays $t\rightarrow c\gamma,~cg,~cZ,~ch$ in the $U(1)_X$SSM. They are certainly very interesting and worth exploring. It also contributes to our understanding of the origin of R-parity and its possible spontaneous violation in supersymmetric models \cite{54,55,56}.

\begin{acknowledgments}
Thanks to Xi Wang, Yi-Tong Wang, Xin-Xin Long for participating in the discussion of the numerical results.
This work is supported by National Natural Science Foundation of China (NNSFC)(No.12075074),
Natural Science Foundation of Hebei Province(A2020201002, A2022201022, A2022201017, A2023201040),
Natural Science Foundation of Hebei Education Department (QN2022173),
Post-graduate's Innovation Fund Project of Hebei University (HBU2024SS042), the youth top-notch talent support program of the Hebei Province.

\end{acknowledgments}

\end{document}